\documentclass[aip,reprint,groupedaddress,showpacs,showkeys,footinbib,citeautoscript]{revtex4-1}
\pdfoutput=1

\newcommand*{\PZT}{Pb(Zr$_{0.2}$Ti$_{0.8}$)O$_3$}
\newcommand*{\SRO}{SrRuO$_3$}
\newcommand*{\STO}{SrTiO$_3$}
\newcommand*{\frefl}[2]{Fig.~\ref{#1}{\color{blue}(#2)}}
\newcommand*{\Vgd}{$V_{gd}$}
\newcommand*{\Vsd}{$V_{sd}$}
\newcommand*{\Isd}{$I_{sd}$}
\newcommand*{\dGrms}{$\delta{}G_{rms}$}
\newcommand*{\lpt}{\tilde{L}_\varphi}

\usepackage{graphicx}

\usepackage{color}

\usepackage[latin1]{inputenc}

\usepackage[bookmarks=false,pdfnewwindow,colorlinks,pdftitle={Understanding polarization vs. charge dynamics effects in ferroelectric-carbon nanotube devices},pdfauthor={Cedric Blaser, Vincent Esposito and Patrycja Paruch}]{hyperref}
\hypersetup{citecolor=blue,linkcolor=blue,urlcolor=blue}

\begin{document}

\title{{\color{blue}Understanding polarization vs. charge dynamics effects in ferroelectric-carbon nanotube devices}}

\author{Cédric Blaser}
\email[Electronic mail: ]{Cedric.Blaser@unige.ch}
\affiliation{\scriptsize DPMC--MaNEP, University of Geneva, 24 Quai Ernest-Ansermet, 1211 Geneva 4, Switzerland}
\author{Vincent Esposito}
\affiliation{\scriptsize DPMC--MaNEP, University of Geneva, 24 Quai Ernest-Ansermet, 1211 Geneva 4, Switzerland}
\author{Patrycja Paruch}
\affiliation{\scriptsize DPMC--MaNEP, University of Geneva, 24 Quai Ernest-Ansermet, 1211 Geneva 4, Switzerland}

\begin{abstract}
(Received 28 March 2013; accepted 16 May 2013; published online 4 June 2013)\\
\\
To optimize the performance of multifunctional carbon nanotube-ferroelectric devices, it is necessary to understand both the polarization and charge dynamics effects on their transconductance. Directly comparing ferroelectric Pb(Zr$_{0.2}$Ti$_{0.8}$)O$_3$ and dielectric SrTiO$_3$ field effect transistors, we show that the two effects strongly compete, with transient charge dynamics initially masking up to 40\% of the ferroelectric field effect. For applications, it is therefore crucial to maximize the quality of the ferroelectric film and the interface with the carbon nanotube to take full advantage of the switchable polarization.\\
\\
\noindent{}\copyright{} \textit{2013 American Institute of Physics. This article may be downloaded for personal use only. Any other use requires prior permission of the author and the American Institute of Physics. The following article appeared in} Appl. Phys. Lett. \textbf{102}, 223503 (2013) \textit{and may be found at} \href{http://link.aip.org/link/doi/10.1063/1.4809596}{http://dx.doi.org/10.1063/1.4809596}
\end{abstract}
\pacs{85.35.Kt, 77.84.-s, 85.50.-n, 77.80.-e, 77.55.fg, 77.55.-g}
\keywords{carbon nanotubes, carbon nanotube field effect transistors, ferroelectric devices, ferroelectric thin films, ferroelectric field effect, ferroelectric materials, transconductance, universal conductance fluctuations}
\maketitle

The exceptional electronic properties of carbon nanotubes (CNTs) have been extensively investigated in field effect transistors, primarily based on Si-SiO$_2$\cite{carbon_nanotube_electronics}. A multifunctional alternative is to combine CNT with ferroelectric gate oxide materials, using the switchable remanent ferroelectric polarization to locally modulate the CNT charge carrier density\cite{sakurai_jjap_06_FEFE_CNT, fu_NL_09_CNT-BTO} at doping levels beyond SiO$_2$ breakdown fields. In such devices, local, nondestructive read-out memory operations have been demonstrated with very low power consumption\cite{fu_nanotech_09_2bit_ferro_CNT}. However, especially at ambient conditions, auxiliary electrochemical phenomena can significantly influence device performance\cite{paruch_apl_08_CNT_ferro, cheah_apl_09_CNTnetwork_FEFE_ntype, nishio_nanotech_09_CNT_PVDF}, in some cases completely screening the ferroelectric polarization. Enhanced by the locally high-intensity electric fields generated directly under a nanoscale electrode such as a CNT, these electrochemical effects can include charge injection and relaxation dynamics, adsorbate effects, introduction and reordering of highly mobile defects such as oxygen vacancies, and even irreversible damage to the ferroelectric material for high gate voltage\cite{kalinin_nano_11_electrochemical_SPM}.

Understanding and controlling the relative importance of these different contributions is crucial for the potential technological integration of ferroelectric-based CNT devices. In particular, since the surface and defect states of both perovskite\cite{sakurai_jjap_06_FEFE_CNT, fu_NL_09_CNT-BTO} and organic\cite{nishio_nanotech_09_CNT_PVDF} ferroelectric gates are often very different from those of SiO$_2$, electrochemical \textit{vs.} polarization effects should be systematically compared in related dielectric and ferroelectric materials.

In this letter, we present such a comparative study of CNT-based field effect devices over a 4K--room temperature range, using ferroelectric \PZT{} (PZT) and dielectric \STO{} (STO) gates, allowing the effects of ferroelectric polarization and charge dynamics to be clearly differentiated. For semiconducting CNT, PZT devices show a dominant ferroelectric field effect, with clockwise, retarding transconductance current-voltage hysteresis. Transient charging/electrochemical effects strongly compete with the ferroelectric modulation of the CNT in dynamic measurements, but decay rapidly at room temperature at zero gate bias, allowing two fully non-volatile memory states. In contrast, STO devices present anticlockwise, advancing transconductance current-voltage hysteresis, and decaying metastable memory states. At low temperatures, increased coercive voltages and largely frozen-out charge dynamics result in qualitatively similar behavior in both types of devices. In metallic CNT, the quasi-static defect landscape leads to reproducible transconductance fluctuations. 

\begin{figure}[t]
\includegraphics{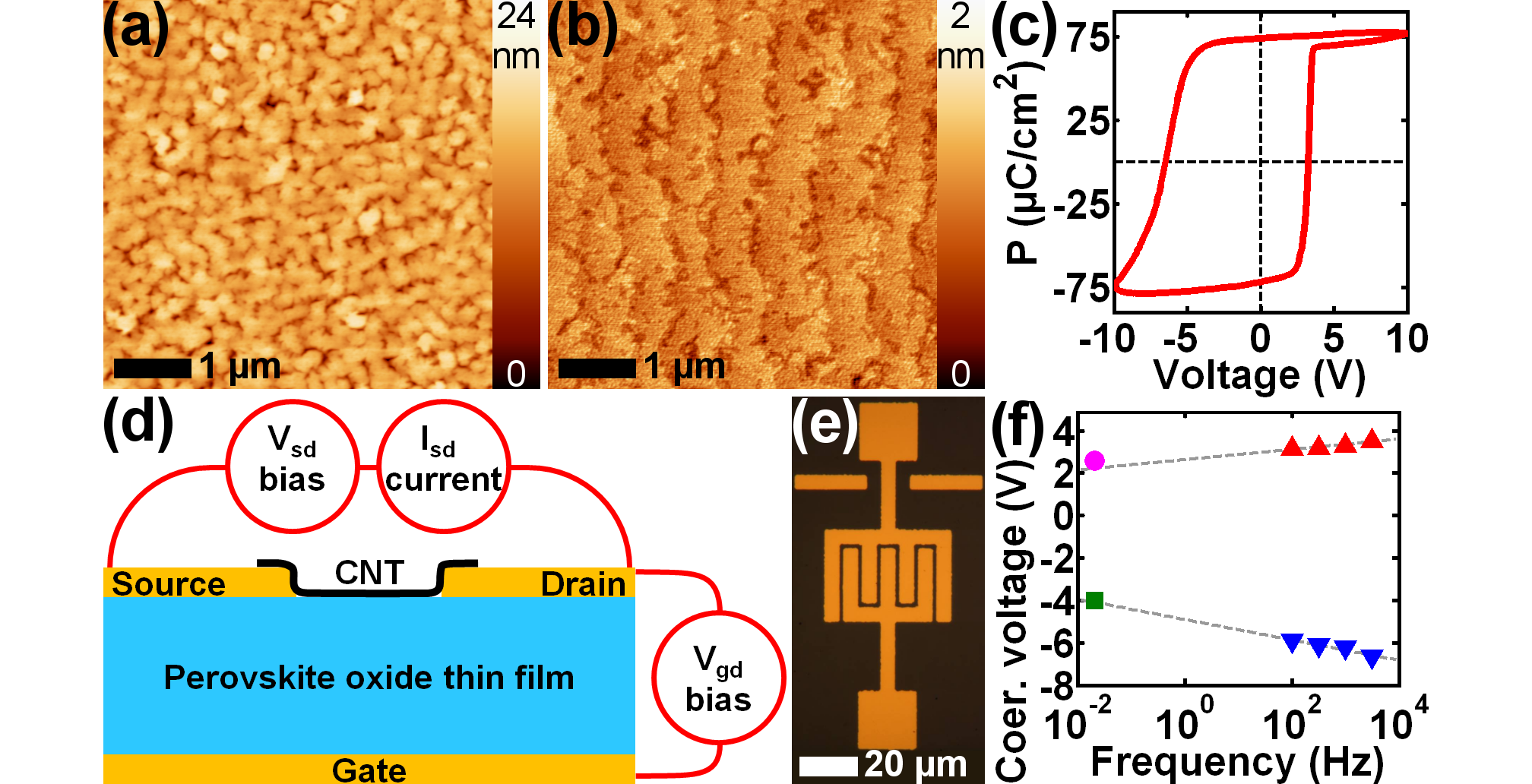}
\caption{\footnotesize \label{fig1}Surface topography of the (a) PZT and (b) STO samples. (c) PZT polarization \textit{vs.} voltage hysteresis at a cycle frequency of 500~Hz. (d) Device schema with \Vsd{} and \Vgd{} bias measured relative to the grounded drain electrode. (e) Optical microscopy image of a device, 1.5~$\mu$m nominal gap. (f) PZT coercive voltages \textit{vs.} cycle frequency. Red up-triangles and blue down-triangles are from polarization \textit{vs.} voltage measurements, magenta disk and green square from the transconductance measurement (\frefl{fig2}{a}). Dashed lines are logarithmic fits of the polarization-voltage hysteresis data.}
\end{figure}

\begin{figure}[b]
\includegraphics{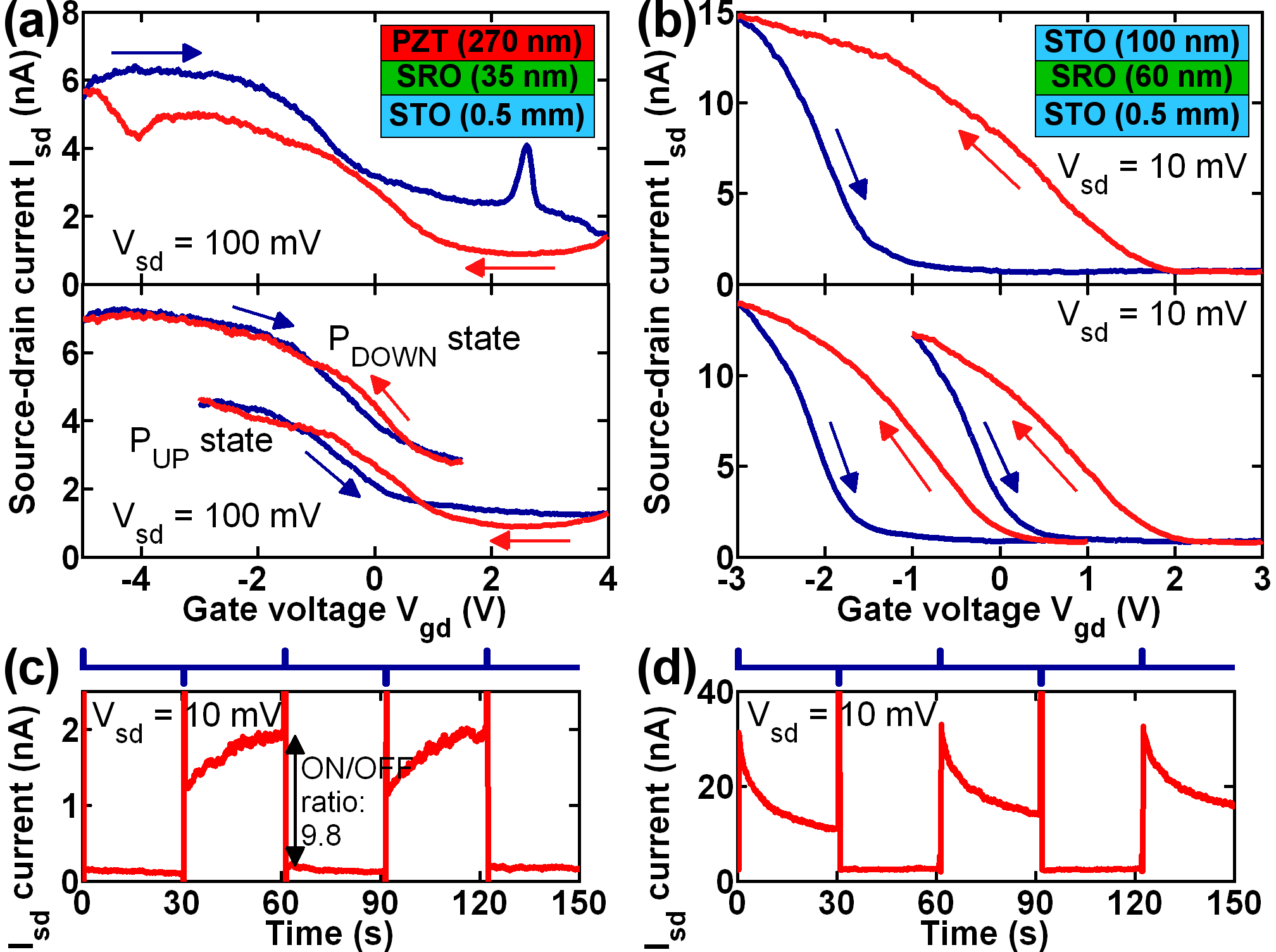}
\caption{\footnotesize \label{fig2}(a) Transconductance measurements at ambient conditions on the PZT sample show clockwise hysteresis, indicating \emph{ferroelectric} field effect. Asymmetric \Vgd{} sweeps avoiding polarization switching show higher \Isd{} with $P_{\rm DOWN}$ than with $P_{\rm UP}$. Inset: PZT sample schema. (b) Transconductance measurement on the STO sample show anticlockwise hysteresis, both for symmetric and asymmetric \Vgd{} sweeps, indicating standard field effect. Inset: STO sample schema. \Isd{} current after short \Vgd{} pulses (represented above the graphs), on the (c) PZT and (d) STO samples, with finite ON current after negative pulses for the PZT sample and decreasing ON current after positive pulses for the STO sample. (a/c and b/d measured on different devices)}
\end{figure}

The samples used in this study were both grown by off-axis radio-frequency magnetron sputtering on conducting \SRO{} (SRO) on (001) single crystal STO (\textit{CrystTec}), with high crystalline and surface quality\cite{gariglio_apl_07_PZT_highTc, zubko_ferro_domains_PTO-STO_2012}. In the PZT (270 nm)/SRO (35 nm) sample (Figs.~\ref{fig1}{\color{blue}(a)} and \ref{fig1}{\color{blue}(c)}, inset of \frefl{fig2}{a}), with 2.4~nm rms roughness, the remanent polarization of $\sim$75~$\mu$C/cm$^2$ is perpendicular to the film plane, and monodomain (`up-polarized') as grown, showing an imprint of $-1.6$~V with coercive voltages of +3.3~V and $-6.5$~V. The absolute values of the coercive voltages decrease with lower frequencies, as shown in \frefl{fig1}{f}. The STO (100 nm)/SRO (60 nm) sample (\frefl{fig1}{b}, inset of \frefl{fig2}{b}) is dielectric, with a rms surface roughness of 0.2~nm. Since both materials are grown by the same technique on similar substrates, similar surface and defect states are expected in the two samples. In particular, Verneuil-grown STO single crystals present relatively abundant dislocations\cite{wang_prl_98_STO_dislocations}, which can propagate into the overlying epitaxial thin films. In addition, the significant chemical activity of these substrates during film growth leads to high densities of oxygen vacancies\cite{yuan_apl_09_Ovac_distribution}. These defects are expected to affect the behavior of CNT in both STO- and PZT-based devices. However, only the PZT-based devices can be expected to show \emph{ferroelectric} field effect. As observed in previous studies, combining CNT and ferroelectrics by direct growth tends to result in deterioration of device properties\cite{paruch_apl_08_CNT_ferro, kawasaki_apl_08_conformal_oxide_CNT}. We therefore dispersed sorted (90\% semi-conducting) single-walled CNT (\textit{NanoIntegris}) by spin-coating directly on the oxide thin film surface from aqueous suspension at a density of 3~CNT/$\mu$m$^2$, after standard photolithographic patterning and electron-beam evaporation of (Cr(5~nm)/Au(50~nm) electrodes (see \frefl{fig1}{e}).

Usable devices, formed wherever at least one CNT connects the source and drain electrodes, are located and measured using BeCu needle probes (\textit{ZN50R-10}) in a cryogenic probe station (\textit{Lake Shore CPX}), operated from ambient conditions down to 4.2~K and 10$^{-7}$~mbar. Variable source-drain (\Vsd) and gate bias (\Vgd) are applied (\textit{NI USB-6259}) while reading the source-drain current (\Isd) after preamplification (\textit{DL Instruments 1211}), as represented in \frefl{fig1}{d}. In agreement with previous observations\cite{sakurai_jjap_06_FEFE_CNT, paruch_apl_08_CNT_ferro, fu_NL_09_CNT-BTO}, non-metallic CNT act as \textit{p}-type semiconductors on perovskite oxide surfaces.

At ambient conditions, transconductance measurements clearly show the competing contributions of ferroelectric field effect and transient auxiliary charge dynamics. In continuous measurements, \Vgd{} is swept repeatedly back and forth, at a rate of 0.36~V/s, at a constant \Vsd{} bias of either 10 or 100~mV, with \Isd{} recorded and averaged over several sweeps, allowing the combined effects of charge dynamics and polarization switching (when present) to be probed. In STO devices, we observe anticlockwise, advancing hysteresis, with a wide 2.2~V window (measured at half-height), as shown in \frefl{fig2}{b}. In PZT devices, we observe clockwise, retarding hysteresis with a highly constricted shape near zero gate bias, leading to a much smaller window of only 0.2~V, as can be seen in \frefl{fig2}{a}. The two distinct polarization switching peaks at 2.6~V and $-4.0$~V are well outside this window.\footnote{See supplementary material at \href{http://link.aip.org/link/doi/10.1063/1.4809596}{http://dx.doi.org/10.1063/1.4809596} for additional transconductance measurements. For the 53 devices measured, the behavior at 0.36~V/s \Vgd{} sweep rate varies from the very narrow window shown to partial or even complete dominance by the transient effects. However as sweep rates were decreased to minimize the influence of transient effects, the hysteretic window widened, highlighting the increasing dominance of the ferroelectric field effect.}
            
When \Vgd{} is swept asymmetrically, avoiding either the 2.6~V or the $-4.0$~V coercive voltage, variations of CNT conductivity in each polarization state can be accessed. As shown in the lower part of \frefl{fig2}{a} for each branch of such asymmetric transconductance measurements, even without polarization switching there is a noticeable \Isd{} increase when \Vgd{} is swept towards more negative values. However, when \Vgd{} remains below the upper coercive voltage (polarization $P_{\rm DOWN}$), \Isd{} values are 50 to 120\% higher than when the polarization is $P_{\rm UP}$. Moreover, although identical \Vgd{} sweep rates are used, there is negligible hysteresis. Meanwhile in STO devices, such asymmetric \Vgd{} sweeps result in qualitatively similar hysteresis to the full \Vgd{} sweeps, although with a reduced window of only 1.0~V, and offset with respect to zero gate bias, as shown in the lower part of \frefl{fig2}{b}.

Finally, by applying 0.5~s \Vgd{} pulses and recording the subsequent \Isd{} evolution at zero gate bias, the non-volatile {\it vs.} transient nature of the ON and OFF states can be probed. In the PZT sample, in agreement with the clockwise transconductance hysteresis of \frefl{fig2}{a}, a positive \Vgd{}~$>$~2.6~V pulse puts the device in the almost non-conducting OFF state (0.15~nA \Isd), while a negative \Vgd{}~$<$~$-4.0$~V pulse switches it ON, as shown in \frefl{fig2}{c}. The initial ON \Isd{} of 1.2~nA increases gradually over $\sim$30~s to 1.9~nA, then remains stable, showing a non-volatile effect of the two opposing polarization states at zero applied field. In the STO sample, as expected from the anti-clockwise transconductance hysteresis, positive \Vgd{} pulses switch ON the device, while negative \Vgd{} pulses switch it OFF (2.5~nA \Isd), as seen in \frefl{fig2}{d}. However, the initially high ON \Isd{} of 32~nA decreases to 14~nA after 30~s, and ON and OFF states reach a common intermediate current value after $\sim$1800~s. 

The results observed in STO devices, with wide anticlockwise hysteresis and an initially high but decaying ON \Isd{} after positive \Vgd{} pulses, are similar to reports on Si/SiO$_2$ field effect transistors,\cite{fuhrer_FET_memory} where the memory effects were attributed to charge injection into metastable states of the dielectric. In addition, surface adsorbates and in particular interfacial water molecules have been shown\cite{kim_nanolett_03_hysteresis_in_CNT_FET} to play a crucial role in the observed dynamic processes leading to such hysteresis. However, while the hold time in the SiO$_2$-based devices exceeded 5000~s under similar conditions, the effects in STO appear to be much more transient, possibly as a result of increased defect densities and higher electron mobility associated with the oxygen vacancies in this material.\cite{tufte_pr_67_mobility_STO}

Comparable defect densities are expected in the PZT sample, and the effects of surface water may even be enhanced by the ferroelectric nature of the sample\cite{hong_apl_10_resistance_hysteresis_graphene_PZT}. The associated charged dynamics strongly compete with the opposing effects of the ferroelectric polarization itself. In epitaxial oxide heterostructures, where wider channel geometries avoid the field focusing and charge injection of CNT-based devices, ferroelectric field effect is characterized by a retarding transconductance hysteresis tracking the positive and negative coercive voltages, and well-defined non-volatile conduction states at zero field, corresponding to the two opposite polarization orientations.\cite{ahn_sci_99_FEFE_hts, takahashi_nat_06_FEFE_sc} In contrast, the present measurements show extremely constricted hysteresis, which remains retarding, but with a very narrow window well away from the actual polarization switching. Another aspect of this competition can be seen in the pulsed measurements at zero gate bias, where as much as 40\% of the final, stable ON state conduction after a negative \Vgd{} pulse is initially masked by transient charge dynamics effects. Likewise, in the asymmetric measurements, although it is always higher in the $P_{\rm DOWN}$ state, significant variation of the CNT conductivity is observed during \Vgd{} sweeps, even though the polarization orientation is not reversed. We note that the magnitude of these variations is well beyond any effective polarization increase due to the enhanced tetragonality of the film \cite{pertsev_prl_98_strain} as a result of its inverse piezoelectric response to the applied voltage.\footnote{For PZT, the $d_{33}$ piezoelectric constant is $\sim$50~pm/V and the polarization $\sim$75~$\mu$C/cm$^2$ for a tetragonality of 1.0475 (out-of-plane lattice constant of 4.148~\AA, in-plane lattice constant of 3.96~\AA). For the asymmetric 7~V voltage sweeps between approximately $-3$ and 4~V, a piezoelectric deformation of 350~pm would therefore be expected. The resulting change in tetragonality of $\sim$0.2\% should not significantly affect the effective polarization.}

Rather, all our measurements point to the key influence of defects, in particular during polarization switching, where the strongest transient charge dynamics effects are observed. In fact, as we have shown previously, the high defect densities present in radio-frequency sputtered ferroelectric films grown on STO substrates significantly modulate polarization switching dynamics and the growth and stability of ferroelectric domain structures\cite{blaser_apl_12_CNT_FE, guyonnet_environmental}. Observing \emph{ferroelectric} field effect in CNT therefore requires high quality materials in which polarization effects will dominate. In ferroelectric BaTiO$_3$ samples with even higher defect densities (resulting from direct CNT growth in reducing conditions), charge injection and relaxation as well as adsorbate effects completely screened out the effects of the polarization, and the devices behaved as standard field effect transistors\cite{paruch_apl_08_CNT_ferro}.

\begin{figure}[t]
\includegraphics{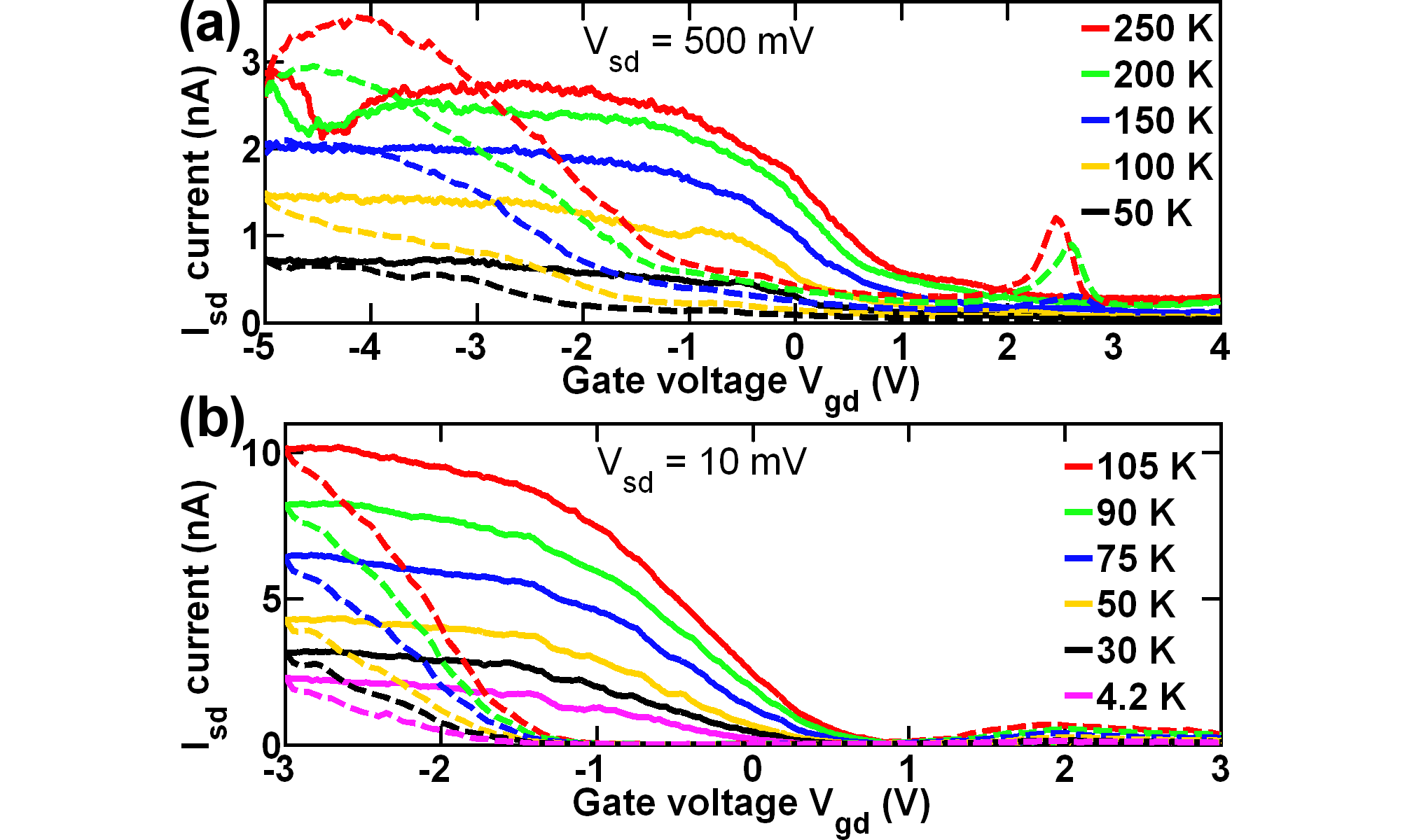}
\caption{\footnotesize \label{fig3}Transconductance hysteresis at 2$\times{}$10$^{-7}$~mbar on the (a) PZT and (b) STO device. In both cases, \Vgd{} down sweeps are shown as solid lines, up sweeps as dashed lines, and measurements at decreasing temperatures present decreasing \Isd{} values. On the PZT device, switching peaks are visible and the leftmost part of the curve shows clockwise hysteresis at 200~K and above. At 100~K and below, no switching peak is visible and the hysteresis is anticlockwise. On the STO device, all measurements show anticlockwise hysteresis, whose amplitude monotonically decreases with temperature.}
\end{figure}

\begin{figure*}[t]
\includegraphics{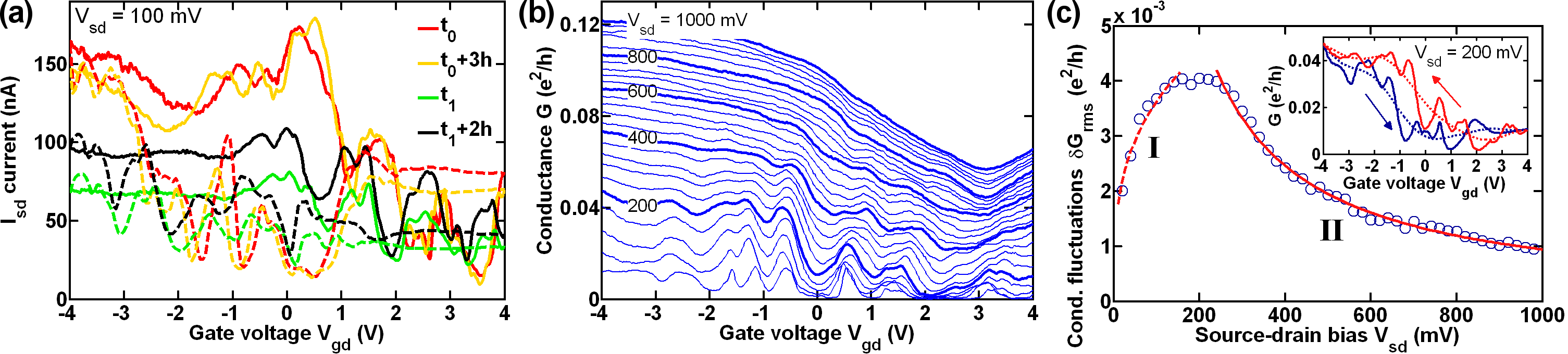}
\caption{\footnotesize \label{fig4}(a) Transconductance measurements at 4.2~K for a STO device containing metallic CNT. \Vgd{} down sweeps are shown as solid lines, up sweeps as dashed lines. $t_1$ is 4 days after $t_0$ and the sample was returned to ambient and cooled down again. The pattern is qualitatively similar during the same experimental thermal cycle, but completely different during subsequent cycles. (b) Conductance of a device containing metallic CNT on a STO sample at 4.2~K as a function of \Vgd{} gate voltage (during down sweeps) for varying \Vsd{} bias. (c) Conductance fluctuations as a function of applied \Vsd{} bias. Inset: extraction of the fluctuations.}
\end{figure*}

To further separate out the effects of the polarization from those of the charge injection and relaxation, the same transconductance measurements were performed at low temperature, where significant slowing of the charge dynamics is expected, and vacuum (2$\times{}$10$^{-7}$~mbar) also removes some of the screening effects of surface adsorbates intrinsically present at ambient conditions. On both the PZT (\frefl{fig3}{a}) and STO samples (\frefl{fig3}{b}), \Isd{} decreases at lower temperatures, as expected for semi-conductors with progressively fewer charge carriers in the conduction band. In addition, in the PZT sample, decreasing temperature also increases the magnitude of the coercive voltages.\cite{piezoelectric_ceramics} Therefore, a constant \Vgd{} range eventually excludes one or both coercive voltage values, and the decreasing temperature can be thought of as ``turning'' the ferroelectric PZT device into one which behaves like a simple dielectric, with no polarization reversal. As seen in \frefl{fig3}{a}, this is exactly what happens as the lower switching peak moves from $-4.4$~V at 250~K to $-4.6$~V at 200~K and finally out of the sweep range at 150~K, whereas the upper switching peak moves from 2.4~V to 2.7~V at the same temperatures. For temperatures below 100~K, only anticlockwise, advancing transconductance hysteresis is observed, with no switching peaks, and the behavior resembles that of the STO-based device in \frefl{fig3}{b}. For the intermediate temperature range between 250~K and 100~K, the competition between charge dynamics and ferroelectric field effect in the PZT device actually leads to a double hysteresis in the transconductance. Whereas at room temperature the ferroelectric field effect is strong enough to just dominate, within the same \Vgd{} sweep range at intermediate temperature, as the coercive voltages gradually increase, the competing effects of charge dynamics become increasingly evident and dominate completely below 100~K.

At the lowest temperatures, where both PZT- and STO-based devices are functionally equivalent, charge dynamics are largely frozen out and so the defect landscape becomes more static. Under these conditions, localization effects and the electronic separation of the CNT into a series of charge puddles can strongly affect the measured transport properties\cite{gao_prb_06_ucf_in_cnts, biercuk_electrical_transport_swcnt}. Since most of the devices contain several CNT, 10\% of which are metallic, we can also access these effects. At room temperature, the presence of any metallic CNT leads to a constant offset of the conductivity versus \Vgd{} bias, resulting in a non-zero \Isd{} current in the OFF state (positive \Vgd{} bias). At low temperature, such devices have a very different behavior. Below 100~K, reproducible fluctuations, whose amplitude increases with decreasing temperature, emerge above the noise level. Their reproducibility is confirmed by multiple sweeps of the same device, including over the span of several hours, as shown in \frefl{fig4}{a}. However, every time the device is returned to ambient conditions and cooled down again, the fluctuation pattern completely changes, indicating the dynamic nature of the disorder potential landscape below the device. The fluctuation pattern is also completely different from one device to another. This phenomenon can be seen on both PZT and STO samples, confirming their functional equivalence at low temperatures.

The uniqueness of the fluctuation pattern between different thermal cycles of the same device, and its repeatability over several hours at constant low temperature are two strong indications of universal conductance fluctuations\cite{[{}][{ [JETP Lett. {\bf 41}, 648 (1985)]}] altshuler_85_conductivity_fluctuations_orig, lee_85_UCF_in_metals}, a disorder-sensitive phenomenon which occurs in the case of phase-coherent transport by the interference of electron waves, which has also been observed in CNT\cite{song_prl_94_electronic_properties_cnt, langer_prl_96_quantum_transport_mwcnt}. As a response to a change in chemical potential or magnetic field, the conductance is expected to fluctuate by the universal value $e^2/h$ as long as the phase-coherence length is at least equal to the channel length. The overall low conductance and the finite temperature of our system may explain the rather small absolute amplitude of the fluctuations observed in \frefl{fig4}{b}.

Following the theoretical work of Larkin and Khmel'nitski\u \i\cite{[{}][{ [Sov. Phys. JETP {\bf 64}, 1075 (1986)]}] larkin_khmelnitskii_original} applied by Terrier \textit{et al.} \cite{terrier_epl_02_quantum_interference} on gold nanowires, we analyze the amplitude of the conductance fluctuations \dGrms{} as a function of the applied \Vsd{} bias, as presented in \frefl{fig4}{c}. The prediction for $V_{sd}\gg{}V_c$, where $eV_c = \hbar{}D/L^2$ is the Thouless energy, $D$ the electron diffusion constant and $L$ the channel length, is an increase of \dGrms{} with voltage $\propto\sqrt{V_{sd}/V_c}$ followed by a decrease as a power law at higher voltages. The increase is explained by the fact that at $V_{sd}\gg{}V_c$, the relevant energy range for the transport subdivides into $N=V_{sd}/V_c$ uncorrelated energy intervals, each contributing to the fluctuations of the current by $\sim(e^2/h)V_c$\cite{terrier_epl_02_quantum_interference}. The amplitude decreases at higher voltages as the consequence of inelastic electron-phonon scattering which suppresses interference phenomena.

The diffusion constant $D$ is not a well-defined quantity, but reported values for CNT range from 50 [Ref.~\onlinecite{song_prl_94_electronic_properties_cnt}] to 900~$cm^2/s$ \cite{schonenberger_apa_interference_mwcnt}. Even with the highest value, our $V_c$ is below 0.03~mV and thus the whole curve of \frefl{fig4}{c} in the regime $V_{sd}\gg{}V_c$. Following Ref.~\onlinecite{terrier_epl_02_quantum_interference}, we fit our data with 
\begin{equation}\delta{}G_{rms}=C\frac{e^2}{h}\sqrt{\frac{V_{sd}}{V_c}}\left(\frac{\lpt(V_{sd})}{L}\right)^2.\label{eq1}\end{equation} 
Because of the device specific constant $C$, only the relative phase-coherence length $\lpt$ can be extracted. In regime I (increasing \dGrms{}), $\lpt\propto{}V_{sd}^{-s}$, where $s=0.09\pm0.04$. In regime II (decreasing \dGrms{}), $\lpt\propto{}V_{sd}^{-p}$, where $p=0.77\pm0.02$. Inserting these values into Eq.~\textcolor{blue}{(\ref{eq1})} results in the dashed and solid lines in \frefl{fig4}{c}. The transition between the two regimes is around 200~mV, which corresponds to the energy for optical phonon emission in single-walled CNT\cite{shi_jap_09_thermal_probing_cnt}, and confirms the role of electron-phonon scattering in the amplitude decrease of the fluctuations. Whereas Ref.~\onlinecite{ludwig_prb_04_conductance_fluctuations} predicts a $1/\sqrt{V_{sd}}$ decay, we observe $\delta{}G_{rms}\propto{}V_{sd}^{-\gamma}$ with $\gamma=1.04\pm0.04$.

In conclusion, we show that the transport properties of semi-conducting CNT on ferroelectric materials are affected by two competing influences: the polarization and the charge dynamics at the film surface. To integrate CNT-ferroelectric devices in applications, it is therefore necessary to optimize the quality of the ferroelectric film and its interface with the CNT to the highest possible level. Realistically, the former is a more accessible goal since, while surfactant and organic adsorbates may be avoidable, under ambient conditions, interfacial water is always present. Since the charge dynamics are transient, these results point to the potentially greater usefulness of such devices for non-volatile zero gate bias applications, once the conductance state is fully stabilized. At lower temperatures, the effects of the strongly varying disorder landscape become even more significant, especially given the increasing coercive voltages for a fixed \Vgd{} range. Eventually, the variations of the disorder landscape dominate completely, leading to characteristic conductance fluctuations in metallic CNT, highlighting the role of electron-phonon scattering at higher \Vsd{} bias.

The authors thank S.~Gariglio for the PZT samples, P.~Zubko and N.~Jecklin for the STO samples, A.~Caviglia, D.~Stornaiuolo and M.~Büttiker for useful discussions, and M.~Lopes and S.~Muller for technical support. This work was funded by the Swiss National Science Foundation through the NCCR MaNEP and Division II grants No. 200021-121750 and 200020-138198.

%

\end{document}